\documentclass[preprint]{aastex}

\input{psfig}

\shorttitle{Grain Aggregation in Supernova Ejecta}
\shortauthors{Deneault}

\begin{document}

%% LaTeX will automatically break titles if they run longer than
%% one line. However, you may use \\ to force a line break if
%% you desire.

\title{Aggregation of SiC-X Grains in Supernova Ejecta}

%% Use \author, \affil, and the \and command to format
%% author and affiliation information.
%% Note that \email has replaced the old \authoremail command
%% from AASTeX v4.0. You can use \email to mark an email address
%% anywhere in the paper, not just in the front matter.
%% As in the title, you can use \\ to force line breaks.

\author{Ethan A.-N. Deneault\altaffilmark{1}} 
\affil{Department of Chemistry and Physics, University of Tampa, Tampa, FL 33606}

\altaffiltext{1}{edeneault@ut.edu}

%% Mark off your abstract in the ``abstract'' environment. In the manuscript
%% style, abstract will output a Received/Accepted line after the
%% title and affiliation information. No date will appear since the author
%% does not have this information. The dates will be filled in by the
%% editorial office after submission.

\begin{abstract}
We present a model for the formation of silicon carbide aggregates within the expanding and cooling supernova 
remnant. SiC-X grains measured in the laboratory at a high spatial resolution have been found to be aggregates 
of smaller crystals which are isotopically 
homogenous. The initial condensation of SiC in the ejecta occurs within a interior dense shell of material
which is created by a reverse shock which rebounds from the core-envelope interface. A subsequent reverse shock
accelerates the grains forward, but the gas drag from the ejecta on the rapidly moving particles limits their travel distance. 
By observing the effects of gas drag on the travel distance of grains, we propose that supernova grain aggregates 
form from material that condensed in a highly localized region, which satisfies the observational evidence of 
isotopic homogeneity in SiC-X grains. 
 
\end{abstract}

%% Keywords should appear after the \end{abstract} command. The uncommented
%% example has been keyed in ApJ style. See the instructions to authors
%% for the journal to which you are submitting your paper to determine
%% what keyword punctuation is appropriate.

\keywords{---supernova remnants ---infrared:stars ---astrochemistry}

%% From the front matter, we move on to the body of the paper.
%% In the first two sections, notice the use of the natbib \citep
%% and \citet commands to identify citations.  The citations are
%% tied to the reference list via symbolic KEYs. The KEY corresponds
%% to the KEY in the \bibitem in the reference list below. We have
%% chosen the first three characters of the first author's name plus
%% the last two numeral of the year of publication as our KEY for
%% each reference.

\section{Introduction}\label{intro}
\indent An important source of interstellar dust is the rapidly expanding and cooling outflows of type II supernovae. Within 
the highly radioactive interior of the ejecta, Si, C and O atoms chemically react in an a nearly hydrogen-free environment, 
forming SiC, CO, and graphite. The most well-studied supernova condensate in the laboratory is the SiC type-X grains 
\citep*{2004oee..symp..300C}. Recent studies of the structure and isotopic makeup 
of these grains \citep*{2004M&PSA..39.5039S,2006M&PSA..41.5333H} show that SiC-X grains are typically comprised of an agregate of smaller 
isotopically homogenous subgrains a a few tenths of a micron to a micron in size. This microstructure indicates that dust coagulation 
processes likely occur within the ejecta. 

\begin{figure}[ht]
\centerline{\psfig{figure=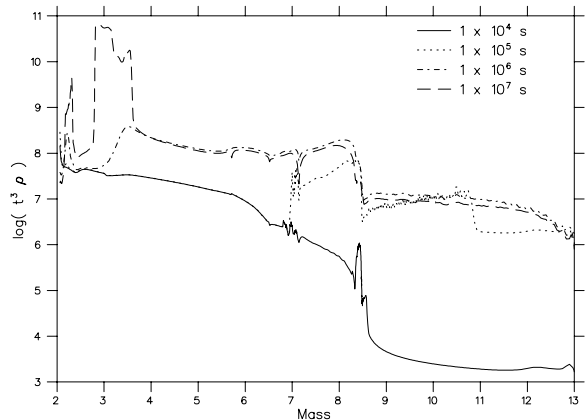,width=3.0in,angle=90}}
\caption{\it Density of the ejecta in mass coordinate plotted for several different times. The large density enchancement 
between 2.7 - 3.6 M$_{\odot}$ contains the SiC condensation region. (This figure replicates Fig.2 from DCH03)}
\label{fig1}
\end{figure}

\indent We follow \citet*[][hereafter DCH03]{2003ApJ...594..312D} in proposing a model for the formation of SiC-X grains in the ejecta, 
in which SiC will condense in the dense ejecta interior, and be accelerated through the overlying ejecta via reverse shocks. 
SiC condenses within the dense, hot interior regions of the supernova ejecta. The reverse shock from the core-envelope interface sets up such a high-density 
region (between 2.7 - 3.6$M_{\odot}$). Figure \ref{fig1} reproduces Fig. 2 from DCH03
to show the large density enhancement which leads to SiC growth. Although the exact condensation kinetics of SiC 
within the hydrogen-free supernova interior is currently unknown, there is indication that strong shocks can drive SiC 
formation \citep*{1989Natur.339..196F}. Within the dense region created by the reverse shock, SiC will only form when the Si/C ratio is 
greater than 10 (DHC03), This condition in satisfied interior to 3.2M$_{\odot}$ and we consider the region 2.7$<m<$3.2M$_{\odot}$ to be 
the primary SiC-condensing region. 

\indent  The SiC grains' motion is coupled to that of the homologously expanding ejecta initially. A second reverse shock (between $10^7$ - $10^9$s after core bounce) 
caused by the interaction between the fast moving ejecta and the surrounding circumstellar material causes these grains 
to be decoupled from the gas, and move forward through the overlying material. 
Recent models of grain survival in ejecta \citep*{2008ApJ...682.1055N, 2007ApJ...666..955N, 2007MNRAS.378..973B} pose that 
upward of 20\% of the dust mass will be destroyed by sputtering in the reverse shock. Smaller condensates are slowed by the drag force 
of the gas and non-thermal sputtering and are eventually destroyed by thermal sputtering while larger condensates a few tenths of 
a micron in size survive roughly intact \citep{2007MNRAS.378..973B}. 

%----------------------------------------------------------------------
%-----------------------------------------------------------------------

\indent Grains that condense farther out in mass coordinate will have a higher velocity than those which condensed closer to the center.
Under this expectation, grains will typically not interact. However, DCH03 posed that larger grains formed in inner mass coordinates 
can overtake smaller grains that formed farther out, because the differential velocity of a grain due to gas drag is proportional to the 
surface area of the grain. By this method, larger grains would act to sweep up smaller grains in their path. The aggregation model of 
DCH03 expected that grains of varying isotopic composition would aggregate together. Current evidence \citep[e.g.][]{2007LPI....38.2321N} suggests otherwise.
Although the nanoscale, intragranular-resolution data on SiC-X grains is limited to a few grains, the available analyses suggest that grain
aggregates are formed from sub-grains that condensed from a single isotopic resevoir. 

\section{Ballistic Motion Through the Ejecta}
\subsection{Reverse Shocks}
\indent The amount of pre-supernova mass loss determines the timescale for the reverse shock to enter the condensation region as well 
as the strength of the reverse shock. If the mass loss is too large, the shockwave will enter the condensation region while SiC growth
is starting. The high temperatures in the shocked gas will stop condensation within the gas, and the small grains that may have already 
condensed will be destroyed by thermal and non-thermal sputtering. A reverse shock that arrives within the condensation region 
after SiC growth has slowed, t $\geq$ 3 years, is more conducive to grain survival. Decoupled from the gas, larger grains are 
more apt to survive non-thermal sputtering, and are not melted by the shock \citep[DCH03,][]{2007ApJ...666..955N}. 

\indent The inital grain velocity depends on where the grain condenses as well as the strength of the reverse shock. The initial grain 
velocity $v_{g0}$ is the difference between the homologous expansion of the gas and the velocity of
the reverse shock. For the purposes of this paper, we take $v_{g0}$ to be a free parameter, and without loss of generality, we choose 
$v_{g0}$ = 500 km s$^{-1}$ to be the ``standard'' initial velocity, following DHC03. 

\subsection{Gas Drag}\label{gasdragsec}
\indent Each grain is decelerated by a drag force 
which is caused by collisions between the grain and the overlying gas as well as non-thermal sputtering from collisions with 
gas particles. Non-thermal sputtering serves to erode the dust grain, slowing its forward motion. In this paper we neglect 
non-thermal sputtering, and thus provide upper limits to travel distances of grains based on gas drag alone. The grain velocity decreases
due to gas drag by \citep*{1979ApJ...231..438D}:    
\begin{equation}
\frac{d v_g}{dt} = -\frac{1}{\rho_g a_g}\left (\frac{3}{2}kT\right )\sum_{i}n_iG_i(s_i)
\end{equation}
where $a_g$ is the grain radius, and $\rho_g$ is the density of the SiC grain, and $n_i$ is the number density of species $i$. The 
function $G_i(s_i)$ is summed over all gas species: 
\begin{equation}
G_i(s_i) = \frac{8 s_i}{3 \sqrt{\pi}}\left (1+\frac{9 \pi}{64}s_i \right )^{1/2}
\end{equation}
and $s_i$ is the atomic speed ratio $s_i = (m_i v_g^2/2 k T)^{1/2}$. In this study, as in DCH03, we use the 25M$_{\odot}$ ejecta model s25 from 
\citet*[][,hereafter WHW]{2002ApJ...576..323R}. This is a one-dimensional ejecta model with solar metallicity, and omits instabilities 
\citep[e.g.][]{1994ApJ...425..814H} that can move inner-ejecta material to higher mass coordinates. The net effect of these instabilities on 
the ejecta is to move the condensation region for SiC grains to a different mass coordinate, but it does not significantly alter the
motion of grains due to reverse shocks, or the condensation chemistry of SiC in these dense clouds. 

\begin{figure}[ht]
\centerline{\psfig{figure=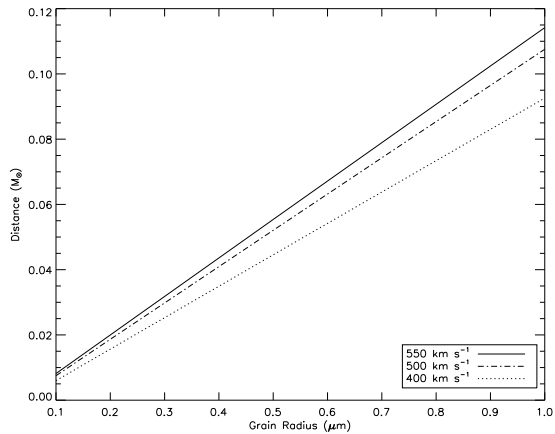,width=3.0in,angle=0}}
\caption{\it Distance in mass coordinate traveled by grains as a function of grain size for different initial velocities. Smaller grains, 
regardless of initial velocity remain near their local region, larger grains travel up to ten times further in mass coordinate.}
\label{distance}
\end{figure}

\indent The effect of the gas drag on the motion of the grain through the overlying ejecta is pronounced. In Figure \ref{distance}, we 
show the distance in mass coordinate traveled by the grain as a function of the grain radius and initial velocity through the ejecta. 
Here we have assumed the grain begins its motion at 3.0$M_{\odot}$ in the s25 model. The total distance traveled is the distance that 
the grain has moved before the gas drag has slowed the grain enough to be comoving with the gas. Smaller grains are effectively and quickly
slowed by the gas drag, slowing to a differential velocity of zero after only a few million seconds, effectlively trapping them within a 
localized region. Larger grains that formed interior to these grains will easily overtake them even though they are moving with a lower
initial velocity.  

%----------------------------------------------------------------------
%-----------------------------------------------------------------------

\subsection{Grain-Grain Interactions}\label{gginteract}

\begin{figure}[ht]
\centerline{\psfig{figure=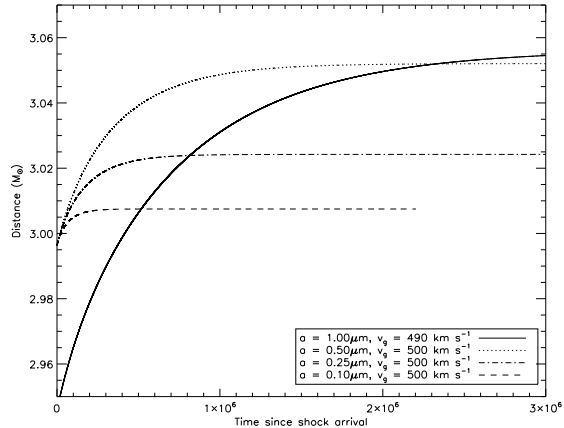,width=3.0in,angle=0}}
\caption{\it Larger grains formed interior to smaller grains can overtake them. The time axis has been scaled such that zero represents 
the time when the shock reached the grain. The grain that formed at 2.95$M_{\odot}$ can interact with grains up to $\sim$0.50$\mu m$ 
in radius that formed a few hundredths of a solar mass farther out.}
\label{interact}
\end{figure}

\indent In Figure \ref{interact}, we see the path of a hypothetical 1 micron grain formed at 2.95$M_{\odot}$. Three other grains that formed in 
the overlying layer at 2.99$M_{\odot}$ are also shown. For clarity, we have not shown grains that may have condensed between those two 
coordinates. The time axis has been scaled such that zero representes the time that the reverse shock
reaches the condensation region of the grain. In the ejecta, the shock reaches the inner zone on the order of 10$^7$s later. Although the 
interior grain has a slower initial velocity, its larger size minimizes the effect of gas drag. In the absence of sputtering or inelastic
collisions, the larger grain can travel approximately 0.1$M_{\odot}$ in mass coordinate. The velocity of grain collisions decreases with 
increasing target size, because smaller grains are effectively smowed by gas drag. 
Low velocity (thus low kinetic energy) impacts have a much higher chance of grains sticking and forming aggregates compared with higher 
velocity impacts, which could shatter the grain. \citep*{1997ApJ...480..647D}. 

%----------------------------------------------------------------------
%-----------------------------------------------------------------------

\begin{figure}[ht]
\centerline{\psfig{figure=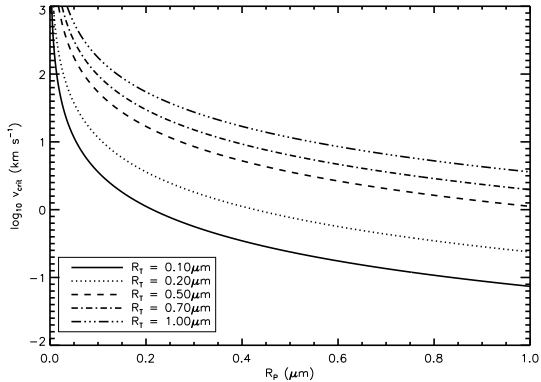,width=3.0in,angle=0}}
\caption{\it Critical velocity for shattering a target grain as a function of radius of the impacting grain ($R_T$). Five target radii are plotted from 
0.1 to 1.0 $\mu m$. Larger SiC grains will shatter smaller grains at even low-speed impacts of a few tenths of a km s$^{-1}$. Only collisions between 
similar sized grains at low velocities will lead to aggregation in lieu of shattering.} 
\label{shatter}
\end{figure}

\indent Although we have not included grain shattering in the calculation, it is worthwhile to consider the effect of grain shattering on the 
evolution of grains within the ejecta. The critical velocity for catastrophic destruction in a grain-grain collision is given by 
\begin{equation}
v_{cat}\propto\left(\frac{R_T^3}{R_P^3}\right)^{9/16}
\end{equation}
Where $R_T$ and $R_P$ are the target and projectile radii respectively \citep*[adapted from][]{1996ApJ...469..740J}. In Figure \ref{shatter}, 
we see that small projectiles, less than 0.1$\mu$m need an extremely large relative velocity to shatter a target grain of any size. However, only 
larger grains can survive catastrophic spallation if the relative velocities between projectile and target are a few km s$^{-1}$. This implies
a very stark picture for the survival of smaller grains. Larger SiC grains which condensed at inner mass coordinate will more likely shatter small 
grains unless the impact velocity is much less than a km s$^{-1}$. Aggregates of similar micron-size crystals have been observed in the laboratory
\citep*[e.g.][]{2007LPI....38.2321N}. These larger grains are more likely to survive impacts with other similar sized crystals at higher velocities. 
Fig. \ref{shatter} shows that a 1$\mu$m impactor will not catastrophically shatter a 1$\mu$m target if the velocity is less than about 4 km s$^{-1}$.

%----------------------------------------------------------------------
%-----------------------------------------------------------------------

\section{Condensate Size Distribution}

\indent Absent any collisions with other grains, a micron sized SiC grain will travel not much more than 0.2$M_{\odot}$ outward before 
gas drag slows its relative forward motion (compared to the ejecta) to zero. This distance represents a strict upper limit to the motion of the grain in the 
ejecta. If the condensation region for SiC grains is found between 2.7$<$m$<$3.2M$_{\odot}$ (as discussed in \S\ref{intro}) that means that SiC grains can
interact with material no further than m = 3.4M$_{\odot}$. In the inner region, near 2.7M$_{\odot}$, the Si/C ratio approaches 10$^4$. 
In this highly Si-saturated region graphite will be readily converted to SiC (DCH03). Due to the much smaller carbon budget in this region however,
it is very likely that the inner condensation zone will produce both a smaller number of condensates as well as smaller-sized condensates
\citep*{2006ApJ...638..234D}.
The specific kinematics of SiC formation as a function of the Si/C ratio are unknown, and we do not endeavor to solve this problem in the current paper. 
As a rule of thumb, we postulate that the size of SiC condensates increases roughly inversely proportional to Si/C ratio. 

\indent As discussed in \S\ref{gasdragsec}, the forward motion of small grains is strongly inhibited by gas drag. In the high-temperature post shock environment, 
thermal sputtering will efficiently destroy small grains. Since the interior condensation region produces only these smaller grains, aggregate formation is 
essential for SiC grain survival in this region. Because a smaller number of initial condensates are formed, however, aggregates from the interior
region will be rare, and comprised primarily of smaller subgrains. Similarly, the Si/C ratio drops below 10 above 3.25M$_{\odot}$, which inhibits the 
growth of SiC. Grains that condensed near this outer edge of the region will not form SiC aggregates with overlying material, and will remain single crystals. 
The central region of the condensation zone, between 2.9 - 3.1M$_{\odot}$, is the most likely place for aggregate formation.  

\section{Isotopic Composition of Grains}

\indent DCH03 proposed that an aggregate of grains would have a noticable isotopic heterogeneity, in that each subgrain of the aggregate
would have a markedly different isotopic composition. Laboratory studies of presolar aggregates show that such isotopic heterogeneity is
rare \citet{2002LPI....33.1297B,2007LPI....38.2321N}, although not unknown. We believe that the observed isotopic homogeneity of supernova aggregates 
indicates that the process of coagulation occurs in a highly localized area of the ejecta. Due to gas drag, the travel distance of a grain is limited. 
It will only have the chance to interact and coagulate with grains that formed nearby. As discussed in \S\ref{gginteract}, large grains, which
decelerated the least by gas drag will have only limited capability of interacting with overlying grains, smaller grains, much less so. 

%----------------------------------------------------------------------
\begin{figure}[ht]
\centerline{\psfig{figure=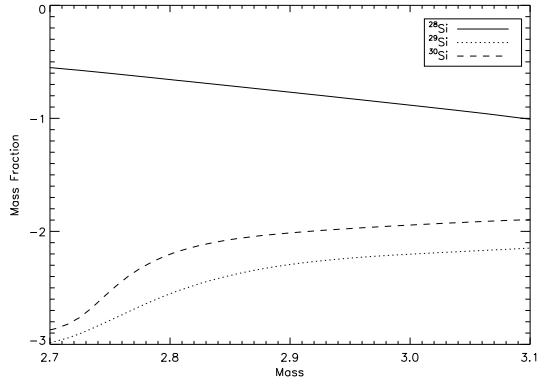,width=3.0in,angle=0}}
\caption{\it Mass fraction of $^{28,29,30}Si$ as a function of mass coordinate in the ejecta. We can see that
the isotopic composition between 2.8 - 3.1M$_{\odot}$ is fairly uniform. Below 2.8M$_{\odot}$, the abundances of $^{29,30}$Si drop by an order of magnitude.
This demonstrates that grains that form near one another will have very similar isotopic composition.} 
\label{siiso}
\end{figure} 
%-----------------------------------------------------------------------

\indent Does this interior region produce the $^{28}$Si excess observed in SiC-X grains? As we can see in Figure \ref{siiso}, the condensation and 
aggregation region between 2.9 - 3.1M$_{\odot}$ does have a substantial $^{28}Si$ excess. This excess varies by a factor of three over the region, 
but is always at least a factor of 10 higher than the abundances of $^{29,30}Si$. If we consider the supernova to be of half-solar metallicity (as per DCH03), 
the $^{28}Si$ abundance will be much higher in this region. The $^{14}N/^{15}N$ ratio, however, is near $1.6\times10^{-2}$ within the region. This is far lower 
than the measured ratio in X grains, which sits between 20-200 \citep*{2000M&PS...35.1157H}. DCH03 proposed that the implantation of $^{14}N$ would occur as the 
grain is propelled through the ejecta, but implantation is insufficient to overcome this problem, especially if the grain's travel is limited by the gas drag. 
Other isotopic ratios, such as $^{26}Al/^{27}Al$, do not match the observed data. The 25$M_{\odot}$ ejecta model that we have used in this paper does not 
consider the possibility of mixing of gases in the ejecta \citet*[][e.g.]{2007ApJ...666.1048Y}. Such ejecta mixing can help to explain the isotopic composition of 
the condensates. 

\section{Ejection into the ISM}

\indent The rapidly decreasing grain velocities in our model effectively retard the total distance that grains can travel relative to the gas. This poses 
a severe problem, as the grains will remain effectively trapped within the expanding ejecta, and not ejected into the ISM. We propose that the grains will 
be ejected into the ISM by a third reverse shock between the outward moving ejecta and the ISM. The time for this reverse shock is 
much longer ($10^9$ - $10^{11}$s) than that of the previous shock due to the circumstellar material. At these late times, the ejecta is very diffuse, and the accelerated 
grains will not be slowed appreciably by the gas drag or by sputtering, and will emerge into the ISM.
 
\section{Conclusion}

\indent In this paper we have presented a model for the aggregation of SiC grains in the supernova interior. In the one-dimensional model, SiC will 
condense within a limited region of the interior where the number density of Si atoms is greater than 10 times that of C. When the reverse shock passes
through the condensation region, the grains are decelerated much less than the surrounding gas, and they ballistically move through the overlying ejecta. 
Gas drag from the ejecta slows these grains down with a rate inversely proportional to the radius of the grain. Larger grains can shatter smaller grains
through impacts, but brains that have low relative velocities can aggregate together. The gas drag highly truncates the distance that a grain can travel, 
thus an aggregate will likely be comprised only of grains with similar isotopic composition. 

\indent We have not set out in this work to show a detailed kinetic description of grain coagulation within the ejecta, but rather to provide a more refined
physical model for that process beyond what was hypothesized in DCH03. The greatest uncertainty in this model is whether grains will actually coalesce in the 
low-speed ballistic collisions that we have presented, and what effect those collisions have on the evolution of the aggregate grain. We have also omitted the 
effects of sputtering and turbulent motion of the ejecta on the grains' travel. Future ongoing work in this area will provide us with a more complete picture of the 
processes of SiC dust aggregation within the ejecta.

% Bibliography Stuff

%% The following command ends your manuscript. LaTeX will ignore any text
%% that appears after it.

\end{document}